\documentclass[useAMS,usenatbib]{mn2e}

\usepackage{graphicx}

\usepackage{txfonts}
\usepackage{multirow}

\def\apj{ApJ}
\def\apjl{ApJ}
\def\araa{ARA\&A}
\def\aap{A\&A}

\def\mnras{MNRAS}
\def\nat{Nature}

\def\apjs{ApJS}

\def\pasp{PASP}
\def\caa{ChA\&A}

\bibpunct{(}{)}{;}{a}{}{,}
\title[Mapping ices in 2D]{Two dimensional ice mapping of molecular cores.\thanks{This work is based on observations with AKARI, a JAXA project with the participation of ESA. 
The James Clerk Maxwell Telescope has historically been operated by the
Joint Astronomy Centre on behalf of the Science and Technology
Facilities Council of the United Kingdom, the National Research Council
of Canada and the Netherlands Organisation for Scientific Research. 
Based on observations carried out under project numbers 088-07, 090-08, and 178-09 with the IRAM 30m Telescope. IRAM is supported by INSU/CNRS (France), MPG (Germany) and IGN (Spain).}}
\author[J.~A. Noble et al.]{J.~A. Noble$^{1}$\thanks{Email: jennifer.noble@u-bordeaux.fr}\thanks{Previous address: Department of Physics, SUPA, University of Strathclyde, John Anderson Building, 107 Rottenrow, Glasgow G4 0NG, Scotland.},
H.~J. Fraser$^{2}$\footnotemark[3], 
K.~M. Pontoppidan$^{3}$,
A.~M. Craigon\footnotemark[3]\\
$^{1}$Institut des Sciences Mol\'{e}culaires (ISM), Universit\'{e} de Bordeaux and CNRS, 351 Cours de la Lib\'{e}ration, F-33405 Talence, France.\\
$^{2}$School of Physical Sciences, The Open University, Walton Hall, Milton Keynes MK7 6AA, United Kingdom.\\
$^{3}$Space Telescope Science Institute, 3700 San Martin Drive, Baltimore, MD 21218, U.S.A.}
\begin{document}

\date{Accepted 1988 December 15. Received 1988 December 14; in original form 1988 October 11}

\pagerange{\pageref{firstpage}--\pageref{lastpage}} \pubyear{2002}

\maketitle

\label{firstpage}

\begin{abstract}
We present maps of the column densities of H$_2$O, CO$_2$, and CO ices towards the molecular cores B~35A, DC~274.2-00.4, BHR~59, and DC~300.7-01.0. These ice maps, probing spatial distances in molecular cores as low as 2200~AU, challenge the traditional hypothesis that the denser the region observed, the more ice is present, providing evidence that the relationships between solid molecular species are more varied than the generic picture we often adopt to model gas-grain chemical processes and explain feedback between solid phase processes and gas phase abundances. We present the first combined solid-gas maps of a single molecular species, based upon observations of both CO ice and gas phase C$^{18}$O towards B~35A, a star-forming dense core in Orion. We conclude that molecular species in the solid phase are powerful tracers of ``small scale'' chemical diversity, prior to the onset of star formation. With a component analysis approach, we can probe the solid phase chemistry of a region at a level of detail greater than that provided by statistical analyses or generic conclusions drawn from single pointing line-of-sight observations alone.
\end{abstract}


\section{Introduction}

The technique of ice mapping was developed to determine the spatial distribution of solid phase molecular species in star-forming regions \citep{Murakawa00,Pontoppidan06}. One question remaining to be answered is whether obtaining information on both the abundances and spatial distribution of solid molecular material simply provides an insight into the global chemical networks in molecular clouds, or whether, additionally, it probes the prevailing physical conditions in the local environment when the ices first formed and can reveal chemical diversity across a star-forming region, thereby illuminating the very small scale feedback between solid and gas phase chemistries.

In general, gas phase molecules are observed in emission, making it comparatively simple to use different chemical species or isotopes to detect molecules and use them to trace astrophysical properties such as density, temperature, and velocity across vast regions of a molecular cloud \citep[e.g.][]{WardThompson07}, or in great detail over very small astronomical scales i.e. tens to hundreds of AU \citep[e.g.][]{Friesen14}. Since ices can only be directly observed as absorption features in the spectrum of a bright background source, mapping their spatial distribution is much more challenging, necessarily relying upon the serendipitous positioning of stars directly behind the cloud to probe individual lines of sight \citep[e.g.][]{Pontoppidan04}. 

One observational approach used to better constrain solid phase chemistry has been to infer that ice must be present in order to explain observed gas phase emission in those cases when molecules with only solid phase formation mechanisms are probed e.g. the detection of gas phase water, generated by non-thermal desorption of water ices deep in a prestellar core \citep{Caselli12}. Such analysis is reliant on gas-grain models, with many data inputs which can be independently changed to influence the output gas abundances for comparison with observations. Similarly, in the study of protoplanetary disks, the gas phase observation of ``complex'' organic species presumed to originate exclusively from the solid phase is used to infer the presence of ice hidden within the disk. The conjecture is that a ``jump'' in the abundance of such molecules in the gas phase is directly relatable to their solid phase abundance through the onset of photo- or thermal-desorption mechanisms \citep[e.g. ][]{Oberg09}. 
Very recently the first observation of a water ice snowline in a protoplanetary disk was made using ALMA, by correlating an increase in continuum dust opacity in the inner disk to the onset of water ice evaporation from cm-sized icy particles migrating from outside the snowline \citep{Cieza16}. The detection of the first branched organic molecule in the ISM, iso-propylcyanide, was also recently reported, and its presence attributed (through astrochemical modelling) to radical addition reactions within the solid phase followed by ice desorption \citep{Belloche14}. 

\begin{table*}
\begin{minipage}[t]{\linewidth}
\caption{Objects observed in this work, and their calculated column densities. }\footnote{Details of the determination of column densities from observational spectra can be found in \citet{Noble13}. Where our fitting routine returned a value of 0 for a component, we have marked these values `bdl' for `below detection limit'. In \citet{Noble13}, Table~3, such cases were marked `0.' Where there was confusion in the spectrum, we provide an upper limit.}
\label{Table1}
\centering
\begin{tabular}{lcccccccccc}
\hline \hline
 Core & ID\footnote{Objects are numbered in order of descending IRAC1 flux. The ID of each object in \citet{Noble13} is shown in parentheses for ease of comparison. Objects marked with an asterisk are YSOs; all other objects are background stars.} & RA              &  Dec            & IRAC1\footnote{IRAC~1 is the flux at 3.6~$\mu$m, extracted from the \emph{Spitzer} c2d catalogues used to identify the objects in the field of view \citep{Evans03}.} 
                                                                                                                                          & N(H$_2$O)                             & N(CO$_2$)                              & N(CO)   & $^{13}$CO$_2$ & CO$_{gg}$ & OCN$^-$\\
(Distance / pc)\footnote{Distances come from \citet{Murdin77,Racca09,lee1999}}  &      &(J2000)         & (J2000)        &mJy        &$\times$~10$^{18}$cm$^{-2}$& $\times$~10$^{17}$cm$^{-2}$ & $\times$~10$^{17}$cm$^{-2}$ & $\times$~10$^{16}$cm$^{-2}$ & $\times$~10$^{17}$cm$^{-2}$ & $\times$~10$^{17}$cm$^{-2}$\\
\hline
\multirow{2}{*}{B~35A}   & 1$^{\ast}$  (3)   & 86.124964 & 9.1492458   &44.60& 3.3~$\pm$~0.1             & 3.7~$\pm$~0.3       & 5.3~$\pm$~3.8 & $<$~3.9 & 1.9~$\pm$~0.4 & bdl\\
                                      & 6$^{\ast}$  (4)   & 86.128544 & 9.1406136    &15.20& 3.1~$\pm$~0.1            & 5.0~$\pm$~0.1       & 3.1~$\pm$~0.9 & $<$~4.8 & 1.1~$\pm$~1.0 & 0.1~$\pm$~0.5\\
  \multirow{2}{*}{(400)}  & 11$^{\ast}$ (2) & 86.122498 & 9.1490002   &3.33& 3.9~$\pm$~0.3                & 4.7~$\pm$~0.2        & 6.7~$\pm$~2.8 & $<$~2.9 & 2.4~$\pm$~0.6 & bdl\\
                                     & 12$^{\ast}$ (5)  & 86.131846 & 9.1494267   &2.44& 1.6~$\pm$~0.3                & 1.8~$\pm$~1.1        & $<$~1.5           & bdl            & bdl                       & bdl\\
\hline
\multirow{2}{*}{DC 274.2-00.4}  & 3$^{\ast}$  (8)   & 142.21379 & -51.616373 &18.20& $<$~0.8          & $<$~2.1                    & $<$~2.3            & bdl           & $<$~0.7            & bdl\\
                                                   & 5$^{\ast}$  (7)   & 142.20929 & -51.610411 &16.90& $<$~1.4          & $<$~1.0                    & bdl                       & bdl           & bdl                       & $<$~0.2\\
       (500)                                    & 13 (6)  & 142.20166 & -51.610514 &2.30& $<$~4.5                         & $<$~1.3                    & $<$~0.6            & bdl           & bdl                      & $<$~0.1  \\
\hline
\multirow{2}{*}{BHR~59}  & 7  (13) & 166.79341 & -62.100358 &9.35& 1.4~$\pm$~0.05                       & 1.5~$\pm$~0.6         & 5.2~$\pm$~9.9  & bdl         & $<$~1.0            & bdl\\
                                        & 8  (14) & 166.79304 & -62.092943 &8.22& 2.6~$\pm$~0.04                      & 3.6~$\pm$~0.5         & 7.2~$\pm$~22.7 & bdl         & bdl                      & bdl\\
            (250)                    & 9  (12) & 166.77597 & -62.101584 &6.57& 0.9~$\pm$~0.03                      & 0.5~$\pm$~0.1         & bdl                         & bdl         & bdl                      & $<$~0.2\\
\hline
\multirow{2}{*}{DC 300.7-01.0}   & 2  (19) & 187.88539 & -63.719644 &35.80& $<$~0.6                    & $<$~0.6                      & $<$~1.6             & bdl          & $<$~3.5            &  bdl            \\
                                                   & 4  (17) & 187.87753 & -63.727139 &18.00& $<$~1.0                     & $<$~0.7                      & $<$~2.6            & bdl           & bdl                       & $<$~1.1 \\
\multirow{2}{*}{(175)}                  & 10 (16)& 187.86635 & -63.722816 &5.14& $<$~0.5                       & $<$~1.0                      & $<$~1.9            & bdl           & bdl                      & bdl\\ 
                                                   & 14 (18)& 187.88413 & -63.725144 &2.24& $<$~1.1                      & $<$~0.9                       & $<$~8.9            & bdl           & $<$~0.8            & bdl       \\
\hline
\end{tabular}
\end{minipage}
\end{table*}  

From a solid phase perspective, there is still much work to do to provide the fundamental data necessary to constrain the origins of these molecules observed in the gas phase, e.g. by determining precise values of ice abundances and compositions, studying ice morphologies, and understanding how solid phase processes are affected by these parameters. Solid phase COM production in astrochemical models is reliant on methanol formation \citep{Garrod08,Drozdovskaya16}, which itself depends on CO ice hydrogenation \citep{Watanabe02}, and by inference, how much CO ice originally formed. This process is known to occur from critical freezeout of CO gas in dense interstellar regions \citep{Jorgensen05}, and CO ice abundances have been shown to vary more signifiantly than any other ice component from star-forming cloud to cloud \citep{Boogert15}. Gas phase mapping can now be conducted on scales of a few AU, probing structure on small astronomical scales within dense molecular cores and even into a disk. The first ``image'' of a CO snowline in a protoplanetary disk was recently constructed from ALMA detections of N$_2$H$^+$ emission (an ion that astrochemical models show increases in abundance as CO freezes out), locating its origin to around 30~AU from the central star \citep{Qi13}. Although such approaches are ideal for locating the onset of CO ice freezeout in the disk, and remain our only spatially resolved probe of ice distributions within protoplanetary disks at the current time, one can not be absolutely certain that ice abundances are consistent (predictable) or even generic over spatial scales of a few 10's to 1000's of AU in star-forming regions. The spatially resolved distributions of CO, H$_2$O, and CO$_2$ ices in dense cores on scales of around a few 1000~AU are reported here for the first time. With these data we aim to address the assumptions inherent in the current observational approach of deriving ice abundances indirectly from gas phase mapping. We generate spatially resolved ice abundance maps and investigate how they compare to gas and dust abundances in the same regions, as well as to the globally-adopted, generic set of interstellar ice abundances \citep{Oberg11}.

The ice absorption spectroscopy method adopted in this study is observationally more challenging than inferring solid phase abundances from gas phase observations. The major advantage of such an approach is that the solid phase material is probed directly; the obvious disadvantages are the limitations on the number of sources and the spatial scale. The traditional approach to ice observations has been to observe single lines of sight in single pointings, then to utilise one-dimensional correlation plots to describe the general trends in ice chemistry over a number of clouds or regions \citep{Gibb04,Boogert08,Boogert13,Noble13}, or histograms to show the range of ice abundances, mixtures, and environments across a statistically significant sample \citep{Oberg11}. The utilisation of a two-dimensional ice mapping approach enables us to test the spread of values in such correlations, probing whether spatial (ice abundance) differences in a core are correlated with localised effects, such as dust evolution, heating, density, turbulence, and radiation field, that are already know to vary considerably over short distances in star-forming regions. Realistically, most of these factors even vary on scales smaller than we are able to probe with current ice maps. 

The first ice maps were reported by \citet{Murakawa00}, who made partial observations of the 3~$\mu$m H$_2$O stretching absorption band towards 50 stars behind the Taurus molecular cloud complex, where the optical depth at 3.1~$\mu$m was found to correlate well with a gas phase C$^{18}$O map, suggesting that H$_2$O ice abundance increases with gas density tracers. The subsequent, more detailed, studies of \citet{Pontoppidan04,Pontoppidan06} highlighted the power of the technique in elucidating ice chemistry and abundance changes on relatively small scales. A map of H$_2$O and CH$_3$OH ice abundances showed that, while H$_2$O ice abundances remain constant over a range of densities, CH$_3$OH ice abundances increase by at least a factor of ten towards the class 0 protostar SMM~4 in the Serpens star forming cloud \citep{Pontoppidan04}. The CH$_3$OH is likely formed from CO frozen out on dust grains at high densities.  This work was followed by H$_2$O, CO$_2$, and CO observations on lines of sight towards five YSOs in the Ophiuchus-F star forming core which, when mapped against distance from the core, implied that CO$_2$ ice formation occurs in two distinct periods: concurrently with H$_2$O at low extinctions, observed in a H$_2$O-rich ice formed on the grain surface; and during CO freezeout, observed in a CO-rich overlayer \citep{Pontoppidan06}. Finally, a study of the column densities of solid CO$_2$ and H$_2$O towards the Cepheus A East star-forming region, was performed using \emph{Spitzer} \citep{Sonnentrucker08}. Uniquely, these ice absorption features were observed, not towards specific IR sources, but against continuum background emission. The results suggested that CO$_2$ and H$_2$O were formed concurrently on dust grains, across the region sampled (much in line with the findings of laboratory experiments on concurrent H$_2$O and CO$_2$ ice formation \citep{Noble11}), while peak abundances were found in regions closest to the embedded protostar. This type of study provides evidence that processing in one part of a star-forming region facilitates a different chemistry in that local area, but not necessarily throughout the surrounding areas. The ice mapping method has even been extended to extragalactic sources; \citet{Yamagishi13} showed that H$_2$O ice is widely distributed in the nearby star-burst galaxy M82, whilst CO$_2$ ice is concentrated near the galactic centre. Whilst again a very different astronomical spatial scale, they also attribute variations in CO$_2$:H$_2$O ice abundance ratios to changes in the ``local'' interstellar environments within the galaxy. Likewise, the same group showed that ice distributions of H$_2$O, CO$_2$, and OCN$^-$ varied significantly across the edge-on galaxy NGC~253, and differed significantly from that of the H$_2$ gas and PAH emission \citep{Yamagishi11}. 

Previously, we reported the detection of ices towards a sample of 22 background stars and eight young stellar objects (YSOs) \citep{Noble13}. The aims of the current work are: using these data, to move from individual spectra to two-dimensional maps of ice column densities for CO, CO$_2$, and H$_2$O ices with spatial resolutions on the scale of a few 1000~AU; to test the ice column densities against the spatial distribution of dust and gas phase CO in the region; and to test the reliability of multi-object, slitless spectroscopy as a technique for ice mapping. We anticipate that the data acquired with AKARI will set the precedent for future spectroscopic observations of this type with JWST and the E-ELT.

   \begin{figure}
   \centering
   \includegraphics[width=0.5\textwidth]{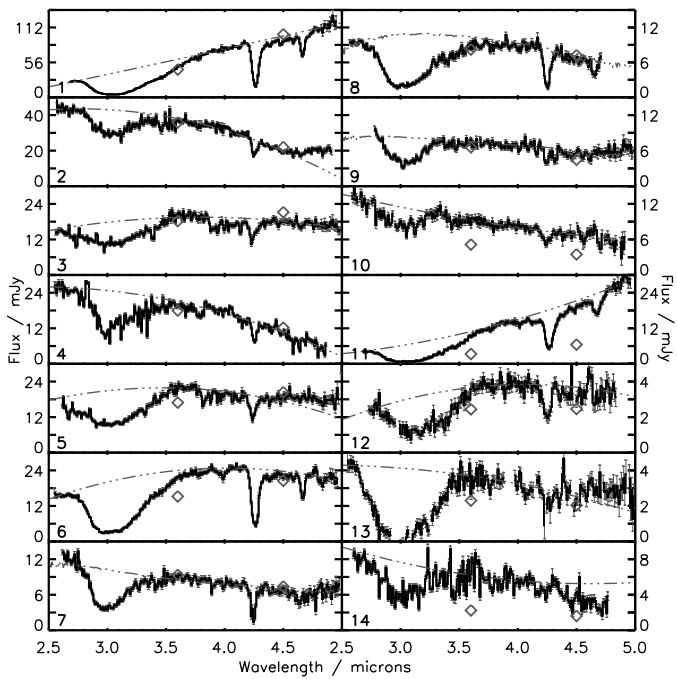}
      \caption{Spectra of the fourteen objects presented in this work. They are numbered as in Table~\ref{Table1}, ordered by decreasing IRAC~1 flux. Overplotted on each spectrum is the fitted baseline (dotted line) and the photometric points IRAC~1 and IRAC~2 (diamonds). Using our extraction technique, we can probe spectra with a flux lower than 10~mJy, as seen in Objects 7--14.}
         \label{Fig1Spectra}
   \end{figure}
%


\section{Observations}

We observed the dense cores of four molecular clouds -- B~35A, DC~274.2-00.4, BHR~59, and DC~300.7-01.0 -- using the NIR spectroscopy mode of the AKARI satellite (AOT IRC04; NG disperser, 2.5~--~5~$\mu$m, 1'~$\times$~1' slit). Data reduction was fully detailed in a previous publication \citep{Noble13}. The aim of the current study was to construct spatially resolved ice maps of H$_2$O, CO$_2$, and CO towards these four molecular cores, chosen from the European Users Open Time Observing Programme IMAPE Programme as they were the only observations from which the spectra of three or more objects could be concurrently extracted from a single AKARI pointing. 

Figure~\ref{Fig1Spectra} shows the extracted spectra, ordered by descending flux at 3.6~$\mu$m (object numbers 1 -- 14). For clarity, the corresponding object numbers used in \citet{Noble13} are given in Table~\ref{Table1} along with target co-ordinates, and the associated ice abundances; targets are identified as either background stars or low-mass young stellar objects (YSOs) according to their attribution in the c2d {\it Spitzer} Legacy programme catalogues \citep{Evans03}. It is interesting to note that ice absorptions are clearly visible by eye in all of these spectra, even at fluxes as low as a few milliJansky (e.g. Objects 12 and 13). Water ice is detected in all 14 spectra. 
CO$_2$ ice is also detected in all objects, but those for which the proximity of other objects in the observed field of view could not be fully corrected for have their extracted column densities given as upper limits in Table~1 (for a more detailed discussion, see \citet{Noble13}). 
CO ice was detected in all but two objects and again around half of these detections include some minor confusion with extraneous light, and so the values are reported as upper limits.
A comprehensive process of baseline subtraction combined with fitting of laboratory spectra was used to extract the ice column densities (again, see \citet{Noble13} for full details). In a few cases, e.g. Objects 2, 5, 13 and 14, the baseline fitted to the full spectrum does not perform as well beyond around 4.5~$\mu$m, so a small additional linear baseline correction was made across the CO absorption feature. The total H$_2$O, CO$_2$, and CO column densities are given in Table~\ref{Table1}. In fact, the total column densities of CO and CO$_2$ ice are the sum of multiple components which reflect the chemical environment surrounding the molecules. The column densities of individual CO and CO$_2$ components, their associated uncertainties, and the component analysis method are reported in \citet{Noble13}. Briefly, in the objects observed in this work, CO$_2$ is present in CO-rich and H$_2$O-rich ices (fitted using CDE-corrected laboratory spectra), while CO is observed in pure (``CO$_{mc}$'') and H$_2$O-rich (``CO$_{rc}$'')  ices (fitted following the \citet{Pontoppidan03} component analysis approach). Additionally, column densities of OCN$^-$, CO gas-grain \citep[CO$_{gg}$][]{Fraser05}, and $^{13}$CO$_2$ were calculated where these spectral features were detected. 

\section{Ice Column Densities}

\begin{figure}
   \centering
   \includegraphics[width=0.5\textwidth]{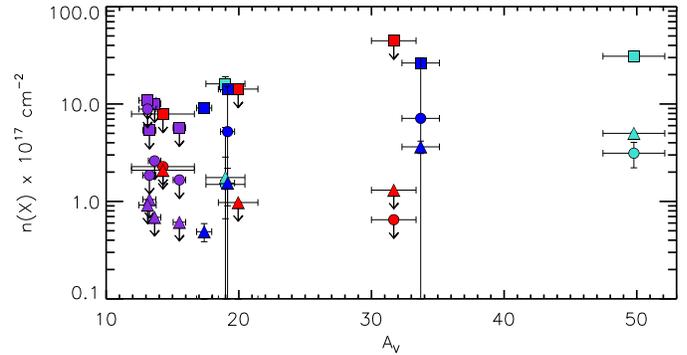}
      \caption{Ice column densities of H$_2$O, CO$_2$, and CO plotted against visual extinction (A$_V$). Ice species are represented by shapes (squares ($\blacksquare$) for H$_2$O, triangles ($\blacktriangle$) for CO$_2$, and circles ($\bullet$) for CO) and cores are identified by different colours (B~35A in cyan, DC~274.2-00.4 in red, BHR~59 in blue, and DC~300.7-01.0 in purple).}
         \label{FigAv}
   \end{figure}

Figure~\ref{FigAv} presents the ice column densities of H$_2$O, CO$_2$, and CO plotted against visual extinction (A$_V$). The data in Figure~\ref{FigAv} is presented two ways: firstly in terms of ice species (squares for H$_2$O, triangles for CO$_2$, and circles for CO), and secondly in terms of cores (B~35A in cyan, DC~274.2-00.4 in red, BHR~59 in blue, and DC~300.7-01.0 in purple). It is known that, generally, as the dust density increases so does the total ice column density, and therefore some correlation is expected between ice column density and A$_V$ \citep[e.g.][]{Whittet88,Whittet07}. We can corroborate this conclusion with Figure~\ref{FigAv}, showing that for our data, broadly speaking, the higher the A$_V$, the more ice is present. Considering the individual ice species, the column densities of CO$_2$ and H$_2$O ice globally increase with A$_V$, whereas the CO column density remains broadly constant. These results are consistent with CO$_2$ and H$_2$O ices forming on the dust surfaces (and therefore increasing in abundance if more dust ({\it i.e.} a greater surface area) is available), but CO freezing out from the gas phase in a process closely linked to the gas density \citep{Pontoppidan06}. The dense cores observed here have extinctions between 10 and 50~A$_V$, so we are not probing the regions of ice formation onset but rather regions with established icy dust grain mantles. 

If we consider the data core by core a more complex picture emerges. In B~35A (cyan), A$_V$ values could only be obtained for two of four sources (target data was extracted from the c2d {\it Spitzer} Legacy programme catalogues \citep{Evans03}), but for the two datapoints available the ice column density increases with dust density. Likewise, in BHR~59 (blue), the ice column densities of H$_2$O and CO$_2$ increase with dust density, and CO ice column densities are approximately constant, as one would expect. However in DC~274.2-00.4 (red) and more particularly in DC~300.7-01.0 (purple) the trends are different. Only the H$_2$O ice column density in DC~274.2-00.4 follows the presumed increase with dust density. The CO$_2$ and CO ice column densities are lower at higher A$_V$. In DC~300.7-01.0, the CO, CO$_2$, and H$_2$O ice column densities decrease with increasing A$_V$, with the trend being most prominent for CO$_2$. This suggests that ice constituents may not reflect the prestellar chemistry across an entire core, but be more heavily influenced by very local conditions. To explain these ice column densities, some ice destruction mechanisms (such as non-thermal desorption) must be dominating, producing effectively lower freezeout rates as A$_V$ increases. 

\begin{figure}
   \centering
   \includegraphics[width=0.5\textwidth]{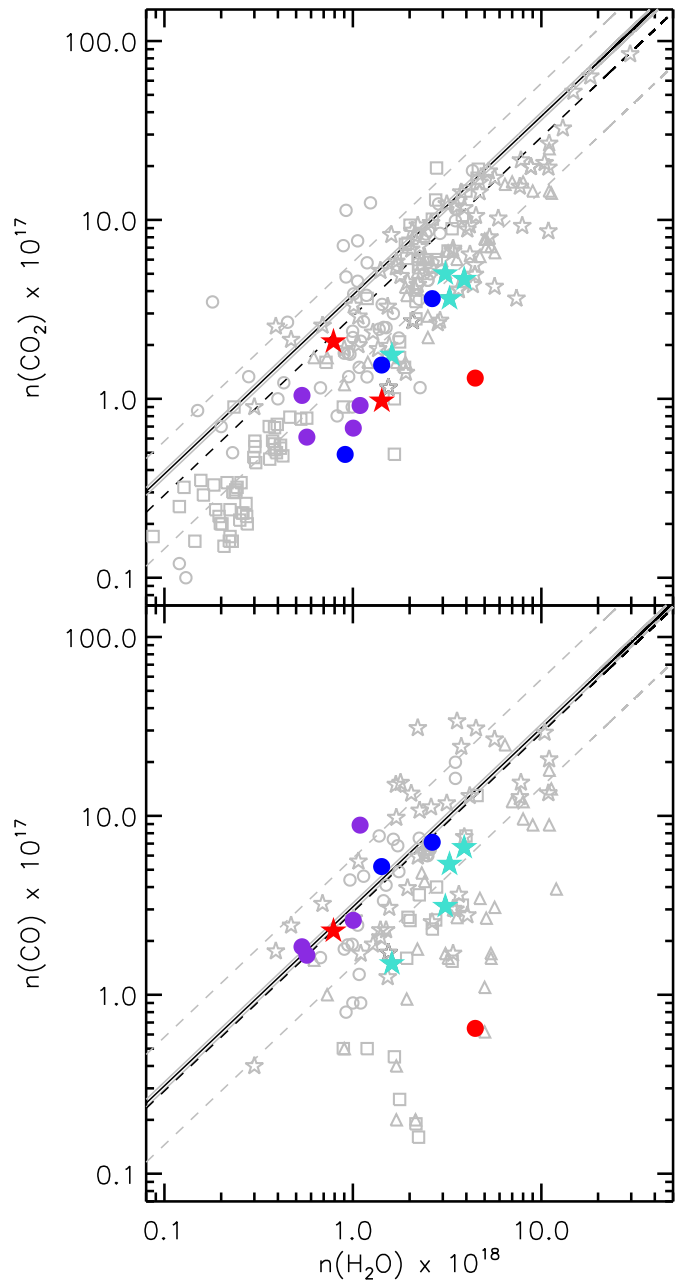}
      \caption{Ice column densities of H$_2$O, CO$_2$, and CO plotted as correlation plots. Object types are plotted as: field stars ($\bullet$), low-mass YSOs ($\star$), high-mass galactic YSOs ($\triangle$), extragalactic sources ($\square$). Literature values are plotted in grey, and the four cores of this work are identified by different colours (B~35A in cyan, DC~274.2-00.4 in red, BHR~59 in blue, and DC~300.7-01.0 in purple). The straight lines represent average ice abundances calculated by \citet{Oberg11} and their derivation is explained in the text.}
         \label{FigCorrelationPlot}
   \end{figure}

We can further probe the adage ``more is more'', {\it e.g.} as the water ice column density increases, so does the CO$_2$ ice column density, by comparing the derived column densities of one ice species to another. This relationship is not entirely true for CO ice, where gas density has been shown to be more critical to the ice freezeout rate \citep{Pontoppidan06}, but holds for H$_2$O and CO$_2$. In Figure~\ref{FigCorrelationPlot}, we plot all published column densities of H$_2$O, CO$_2$, and CO ices (of which we are aware, per single source, excluding upper-limits) in two correlation plots: H$_2$O versus CO and H$_2$O versus CO$_2$ \citep{Gerakines99,Nummelin01,Gibb04,vanBroekhuizen04,Bergin05,Knez05,Whittet07,Pontoppidan08,Whittet09,Zasowski09,Oliveira10,Shimonishi10,Boogert11,Chiar11,Noble13,Oliveira13,Poteet13,Boogert13,Whittet13,Yamagishi15}. The data exploited in this work, which originate from background sources and low-mass YSOs, are shown in colour on these plots and entirely overlap with existing literature values. We will revisit the specific core-by-core distribution of our data below. It is clear that, although the derived column densities of each species span between two and three orders of magnitude, a general correlation is apparent in each plot.

The limitation of traditional column density correlation plots is that they give us no information about the astronomical source of each data point, or the extent to which ice column densities change on scales of up to a few thousand AU {\it i.e.} which data point is associated with which particular line of sight, nor which objects lie within, or probe, the same cloud or core. Clearly, the correlations contain a significant spread of values (in the anti-correlation direction), even when plotted on log-log scales, which leads to the question: how can we account for this diversity? One might suspect ice column density of varying consistently with source type. The plots in Figure~\ref{FigCorrelationPlot} include both extragalactic data (squares) and galactic data, the latter divided into low-mass YSOs (stars), high-mass YSOs (triangles) and background sources (circles). Extragalactic values are typically closer to those found in high-mass YSOs in our own galaxy \citep{Oliveira10,Shimonishi10,Yamagishi15}. However data from galactic low-mass YSOs, background stars and high-mass YSOs are well spread across Figure~\ref{FigCorrelationPlot} with no evident source-dependent trends. This spread of values was rationalised in a detailed statistical analysis by \citet{Oberg11} of ice observations encompassing 63 low-mass YSOs, eight high-mass YSOs and a handful of background (galactic) sources (without access to the more recent field star surveys of \citet{Noble13,Boogert13}). \citet{Oberg11} derived the median generic ice abundances of (amongst other solid phase species) H$_2$O, CO, and CO$_2$, giving typical values of 100:29:29 (in low-mass YSOs), 100:31:38 (in cloud cores {\it i.e.} towards background stars), and 100:13:13 (in high-mass YSOs). These generic ratios are superimposed on Figure~\ref{FigCorrelationPlot} as black lines: solid for background sources and dashed for low-mass YSOs, which for these three ice species have remarkably similar ratios. However, they also noted that (in their dataset) the actual abundance values towards low-mass YSOs could vary for CO and CO$_2$ by up to a factor of two (although towards background stars the variation was only a few percent). Figure~\ref{FigCorrelationPlot} shows that the published ice column density data from H$_2$O, CO, and CO$_2$ ices are mostly encompassed by the limits calculated by \citet{Oberg11} (shown as grey solid lines for background sources and grey dashed lines for YSOs). However, we include more data points than \citet{Oberg11} and observe that the spread of values is even broader (in this case, for H$_2$O:CO$_2$ we plot 250 objects (comprising 62 background stars, 80 low-mass stars, 31 high-mass stars and 76 extragalactic sources) and for H$_2$O:CO we plot 127 objects (comprising 33 background stars, 47 low-mass stars, 30 high-mass stars and 17 extragalactic sources)). This large data spread is also observed within the small subset of our own data points (plotted in colour).

Clearly then, source type alone can not account for the diversity of ice column densities observed. An alternative explanation is that from one cloud to another one occupies a slightly different region of parameter space on the correlation plot, which somehow reflects the differences on astronomically ``regional'' scales. On a cloud-by-cloud basis, variations in ice formation and processing in prestellar phases would be carried forward to the protostellar phases, meaning that different cores within the same cloud would demonstrate similar ice abundances. However, it is also possible that a more local core-to-core variation is seen within one cloud. In Figure~\ref{FigCorrelationPlot}, the ice column densitites from each of our cores are highlighted in cyan (B~35a), red (DC~274.2-00.4), blue (BHR~59), or purple (DC~300.7-01.0). Since H$_2$O and CO$_2$ ices were detected in all 14 objects in this data set, the column density correlation plot for CO$_2$ versus H$_2$O contains a full compliment of data, whereas the CO versus H$_2$O correlations exclude two objects (where no CO ice was detected).  Although the statistics are relatively poor (because the field of view is only 1 arcminute square and therefore contains a maximum of four non-confused objects in each case), of the four cores presented here, two  -- B~35A and BHR~59 -- adhere globally to the ``more is more'' paradigm while the other two -- DC~300.7-01.0 and DC~274.2-00.4 -- manifestly do not. It is clear from Figure~\ref{FigCorrelationPlot}, that the correlation between the CO$_2$ and H$_2$O column densities in the four YSOs in the B~35A field of view (highlighted in cyan) lie almost entirely on a straight line. However, the lines of sight towards background stars in BHR~59 (blue) already deviate slightly from a perfect correlation, while the background stars in DC~300.7-01.0 (purple) and the combination of background stars and YSOs in DC~274.2-00.4 (red) clearly show no correlation, and potentially even an anti-correlation, between the H$_2$O and CO$_2$ column densities. Likewise, even though the correlation between H$_2$O and CO column densities is generally less robust than that between CO$_2$ and H$_2$O, in B~35A and BHR~59 the ``more is more'' adage still seems to hold, whereas again in DC~300.7-01.0 and DC~274.2-00.4 it does not. Realistically there is no clear trend in ice abundances, even within those cores that contain only background sources, so it appears that ice column densities must be reliant on much more local influences than cloud-to-cloud or even core-to-core variations. This conclusion is entirely consistent with the statistical Kruskal-Wallis analysis of \citet{Oberg11} and an earlier principal component analysis by \citet{Oberg09}. 

Individual variations in ice abundances must therefore be attributable to very localised conditions during both the prestellar formation phase of the ice and its subsequent processing during protostellar evolution, even within individual cores, as suggested by the Monte Carlo simulations of \citet{Makrymallis14}. For the remainder of this paper we consider three distinct phases of ice formation and processing \citep[e.g.][]{Pontoppidan06,Burke10,Oberg11}, and discuss our ice data in this context. Of specific relevance to the three ice species discussed here:
\begin{itemize}
\item firstly, ices form such that the CO$_2$:H$_2$O column density ratio is determined in the ``early'' stages of ice formation, in prestellar cores, when the CO$_2$ ice forms in competition with H$_2$O ice, driven by hydrogenation reactions, thereby justifying why these ices generically follow the ``more is more'' mantra. Such mechanisms are corroborated by laboratory experiments which quantify the relative rates of CO$_2$ and H$_2$O formation \citep[e.g.][]{Oba10,Noble11} and by models which exploit such empirical data \citep[e.g.][]{Reboussin14,Walsh14,Esplunges16}. Devations from ``more is more'' must point towards specific localised conditions which accellerate or decellerate ice formation with increasing A$_V$.
\item secondly, ices evolve such that the CO:H$_2$O column density ratios are constrained mostly during the ``late'' prestellar phases, where CO ice freezes out and ice composition is driven by variations in CO freezeout rate, combined with a CO-based chemistry. This has been shown in the first multi-molecular 2D ice maps of the Ophiuchus-F core \citep{Pontoppidan06} and the conclusions of the statistical analysis of \citet{Oberg11}.
\item finally, the CO:H$_2$O column density ratio varies again in protostellar environments, where ice chemistry is subject to thermal and non-thermal processing, leading (of particular relevance here) to CO ice desorption, ice mixing and/or segregation between ice constituents, and processing to form more complex chemical species.
\end{itemize}
This highlights the need to be able to map in at least two dimensions the positions and ice abundances in objects concurrently, and to be able to compare such maps with other observational constraints on the prevailing physical conditions in the ISM. For the remainder of this paper we focus on this challenge.

%

\section{Ice maps}\label{SecIceMap}

\begin{figure*}
\includegraphics[width=\textwidth,clip=true, trim=0 40 0 0]{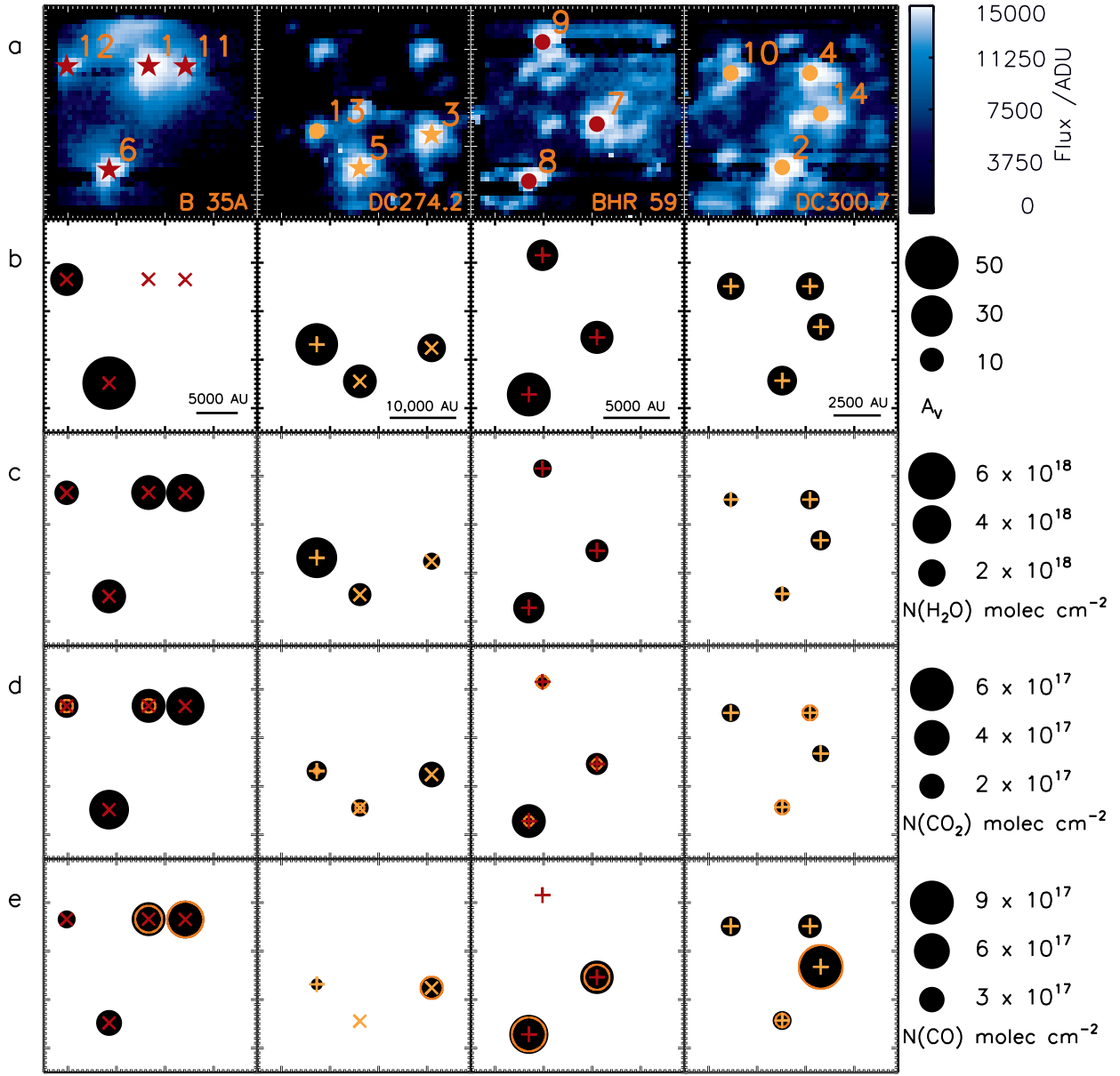}
  \caption{Ice maps of molecular cores B~35A, DC~274.2-00.4, BHR~59, DC~300.7-01.0. Objects are numbered from highest to lowest IRAC~1 flux, as in Table~\ref{Table1}. Tickmarks represent the pixels of the detector, where 1~pixel~$\simeq$~1.46~arcseconds. In a) objects are marked with $\star$ for YSOs, or $\bullet$ for background stars, while in b~--~e these symbols are replaced by $\times$ and $+$, respectively, for clarity (object positions marked in yellow yielded only upper limits on column densities). For all maps, the area of the black circle corresponds to the quantity being mapped, according to the scale on the right hand side. The maps presented are as follows: a) Imaging frame from AKARI (with object positions shifted to reflect the slight shift in AKARI fields of view, see \citet{Noble13}). b) A$_V$, taken from the {\it Spitzer} c2d catalogues \citep{Evans09}. Objects~1~\&~11 have no calculated A$_V$ values in the catalogues. c) Column density of H$_2$O. d) Total column density of CO$_2$ is plotted in black. Additionally, the column density of the component CO$_2$ present in a H$_2$O environment is overplotted as orange circles. e) Total column density of CO is plotted in black. Additionally, the column density of the component CO present in a H$_2$O environment (CO$_{rc}$) is overplotted as orange circles. Objects~5~\&~9 have no CO ice.\label{Fig2Maps}}
\end{figure*}

Maps of the calculated ice column densities are presented in Figure~\ref{Fig2Maps}. The AKARI imaging fields of view at 3.2~$\mu$m are shown in Figure~\ref{Fig2Maps}a. Adopting the c2d definitions \citep{Evans09}, the BHR~59 and DC~300.7-01.0 fields of view contain only background star observations (marked $\bullet$), and so are likely to trace prestellar regions, while the map of B~35A contains only data from embedded objects (marked $\star$) and DC~274.2-00.4 contains both. The two closest objects which could be extracted from the AKARI field of view using our reduction pipeline are Objects~4~\&~14. As DC~300.7-01.0 lies at a distance of 175~parsecs \citep{lee1999}, the resolution between these objects represents $\sim$~2200~AU, and thus the map probes chemistry over small spatial distributions (but still a significantly larger scale than gas phase observations). Figure~\ref{Fig2Maps}b shows the A$_V$ towards each object plotted as bubble plots, the values being extracted from the c2d catalogues. Objects~1~\&~11 have no calculated A$_V$ values, but based on gas phase observations of this cloud, it is estimated that the values will be roughly equivalent to the other two objects in the field of view \citep{Li2007}. Visual extinction is correlated to the dust abundance in the cloud, and as pointings were centred on the most dense region of each core, we are likely probing the highest visual extinctions possible. 

H$_2$O ice maps (Figure~\ref{Fig2Maps}c) show column densities of H$_2$O which are closely related to the A$_V$ values in the panels above. The ice maps are testing the spread in a 1D correlation plot of H$_2$O and A$_V$. Our data agree well with previous studies \citep[e.g.][]{Whittet88} that A$_V$ and N(H$_2$O) are closely correlated. Although the relationship is consistent within the clouds, it is not possible to predict the column density of H$_2$O given the A$_V$, nor vice-versa. For example, in DC~300.7-01.0, the A$_V$ is approximately constant towards all lines of sight, but the column density of H$_2$O varies by a factor of two over a small scale. There must, therefore, be other physical and astronomical influences to consider in the formation and spatial distribution of ice in clouds, and ice mapping provides a powerful tool for investigating them. 

The information in an ice map illustrates the link between chemical reagents and products on a spatial scale, and CO$_2$ and CO are a good test of this. In Figure~\ref{Fig2Maps}d, the maps of total CO$_2$ ice column density show a general agreement with the H$_2$O ice maps. This is unsurprising, as previous studies have already established the existence of a correlation between the abundances of H$_2$O and CO$_2$ in interstellar ices \citep[e.g.][]{Whittet07,Pontoppidan08,Noble13}. A notable exception to this pattern is Object~3, where the column density of CO$_2$ appears to be high in relation to that of H$_2$O. Thus, in clouds such as B~35A, CO$_2$ appears to be a tracer of H$_2$O, while in DC~300.7-01.0 this does not seem to be the case. However, given that the values calculated in the former are upper limits only, it is not possible to draw any firm conclusions based upon this single aberration. 
CO ice maps are shown in Figure~\ref{Fig2Maps}e and are in general agreement with both the A$_V$ and H$_2$O maps. All of the lines of sight are above the critical extinction threshold for CO freezeout, and so the column densities of CO are expected to be approximately 20~--~30~\%~H$_2$O \citep{Chiar95,Boogert04}. However, we would expect to see discrepancies in the CO distribution as it begins to freeze out. Towards Objects~5 and 9 there is no CO ice, but this is a reasonable result when the spectra in Figure~\ref{Fig1Spectra} are considered. There is little or no CO absorption band visible in these spectra, suggesting that, within the detector limits, there is no CO ice towards these lines of sight. A potential explanation for this is that heating in these isolated regions has caused CO to desorb into the gas phase.

In general, observations to date have shown that ice abundances toward background stars and protostars are largely similar, except for a lack of features associated with ice heating (for example pure CO$_2$ ice) toward background stars. In particular, there is no evidence for a different range of CH$_3$OH and CO$_2$ ice abundances. Additionally, the CO$_2$:H$_2$O ratio, and CH$_4$ and NH$_3$ abundances vary little with respect to H$_2$O, suggesting co-formation of these ices. In contrast, the CO:H$_2$O and CO$_2$:CO ratios, as well as the abundances of OCN$^−$ and CH$_3$OH vary by factors of two to three with respect to H$_2$O, indicative of a separate formation pathway from H$_2$O ice. Pure CO and CO$_2$ ice abundances vary even more, consistent with their sensitivity to protostellar heating \citep{Oberg11,Boogert15}. Interestingly, across all ice species mapped in Figure~\ref{Fig2Maps}c, d~\&~e, there is no real trend in the column densities derived for either background stars or for YSOs. In these data, the highest column density of H$_2$O is observed towards a background star (Object~13). However, it must be noted that this value is an upper limit, and the next three highest H$_2$O column densities are seen towards YSOs in B~35A. For CO, the two highest column densities are seen towards background stars (Objects 8 and 14), only one of which (14) is an upper limit. For CO$_2$, the three highest column densities are again seen towards YSOs in B~35A, which fits better with the standard scenario of ice formation. 

When considering the component analysis of CO$_2$ and CO (as shown in Figure~\ref{Fig2Maps}d,e by the overplotting of orange circles denoting water-rich environments), it is interesting to note that in the mapped data there is no obvious pattern in any of the fields of view. For example, towards B~35A, the CO$_2$ towards Objects~6 and 11 is entirely in a CO-rich ice (the H$_2$O-rich component, denoted in these maps by orange circles, being competely absent), and towards Objects~1 and 12, the majority is in a CO-rich ice. For CO, Objects~6 and 12 are exclusively CO in a pure CO ice, but Object~11 has predominantly CO in a H$_2$O-rich ice, and Object~1 is more equally mixed. Thus, for Object 11, (almost) all CO is in a H$_2$O-rich environment, rather than pure CO, but the CO$_2$ is exclusively in a CO-rich ice. Is it possible that CO$_2$ is trapped in CO pockets in a H$_2$O ice? In Object 6, however, all the CO and CO$_2$ is in a CO-rich environment. Towards this line of sight, did CO$_2$ form exclusively from a pure CO ice? Or is the ice porous, and diffusion of CO and/or CO$_2$ has occurred on a large scale? Is it possible that the CO freezeout occurred extremely rapidly in this line of sight? Furthermore, CO and CO$_2$ ice components (e.g. in water) can vary by factors of at least an order of magnitude from the ``median'' values derived. These results somewhat challenge the standard model of ice formation, which is based upon the freeze out of a CO layer on top of a mixed CO$_2$:H$_2$O layer. Considering the other fields of view, it seems that there are also some other objects with non-standard column densities. Towards Object 3 in DC~274.2-00.4, all CO$_2$ is in a CO-rich ice, while all CO is in a H$_2$O-rich ice (which is similar to Object 11 in B~35A). Towards Object 14 in DC~300.7-01.0, there is an anomalously high CO column density, while the little CO$_2$ which is present is again exclusively in a CO-rich ice. These data suggest that even within a single cloud, the chemistry can be different towards different lines of sight and that making generalisations might not be a reasonable approach to take. The relationships established by the component analysis comparisons in 1D correlation plots become more complex still when spatial distribution is considered. 

If it is assumed that some CO$_2$ formation occurs at early times, concurrent with H$_2$O, it would be reasonable to assume that some CO$_2$ in a H$_2$O-rich component would be visible towards all lines of sight. The component analysis in Figure~\ref{Fig2Maps}d shows that no CO$_2$ in H$_2$O is observed towards Objects~6, 8, 10, 11~\&~14. This is most likely due to a limitation in the detection method, with very small abundances of CO$_2$ in fact being present on these lines of sight, but requires observations at higher sensitivity and resolution for clarification. As was discussed earlier, most of the CO$_2$ in a H$_2$O-rich environment is likely formed at later times, so correlation should be seen between CO$_{rc}$ and CO$_2$ in H$_2$O. This is generally the case in B~35A and BHR~59, but there are anomalies. For example, Object 9, which has no CO ice, has almost all its CO$_2$ ice in a H$_2$O-rich phase. This could suggest that all of the CO has desorbed or has reacted to form CO$_2$ in the H$_2$O-rich ice, as seen in the ice models of \citet{Pontoppidan03,Pontoppidan08}, and could be evidence of heating, UV processing, or the age of Object~9.

Considering minor components which were observed towards some lines of sight, but are not mapped here, Objects 3, 4, 5, 6~\&~9 were all found to have OCN$^-$, which is a tracer of UV or temperature processing. This corresponds well to the assertions made about the non-detection of CO towards Objects~5~\&~9. Object~6 has the highest (available) A$_V$ of any of the mapped objects, and thus is the most highly embedded. It may be that the observations are more sensitive to OCN$^-$ at higher extinctions. Object~5 has low CO, CO$_2$ column densities, and an OCN$^-$ detection, which could suggest that this region of DC~274.2-00.4 is heated. There is some evidence of the presence of CO$_{gg}$ (gas-grain) towards all clouds, but not towards all lines of sight. As CO$_{gg}$ is believed to be a component that corresponds to CO interacting directly with bare dust grain surfaces, it could trace a CO component that can form CO$_2$ at very early times in a cloud, before there is accumulation of an icy mantle on grains. In B~35A, Objects~1, 6~\&~11 all have CO$_{gg}$ components, and it is also present towards Objects~2, 7, 13~\&~14. As this is a mixture of background stars and YSOs, with varying extinctions, a correlating factor is not immediately evident. Further observations would provide more insight into this issue. 

These data highlight the potential importance of the ice mapping technique in extracting detailed chemistry from prestellar regions, and represent an important first step in our understanding. They strongly suggests that chemistry in interstellar regions could be very local, changing even from object to object within one cloud. Ultimately, higher resolution spectra over a wider spectral range (e.g. combined NIRSpec/MIRI observations using JWST) are required to fully investigate the environments in which ice and gas phase molecular species are present towards these lines of sight. Gas-grain coupled astrochemical models can be used to predict ice abundances, either at the ``end'' of a prestellar phase \citep{Makrymallis14}, or as a function of time during protostellar evolution \citep{Viti04}, and an ideal observational constraint for comparison would be predictions of ice abundances for all species, even if observations only existed for one or two of them.

\section{Combined solid-gas maps}\label{SecGasMap}

\begin{figure*}
\includegraphics[width=\textwidth]{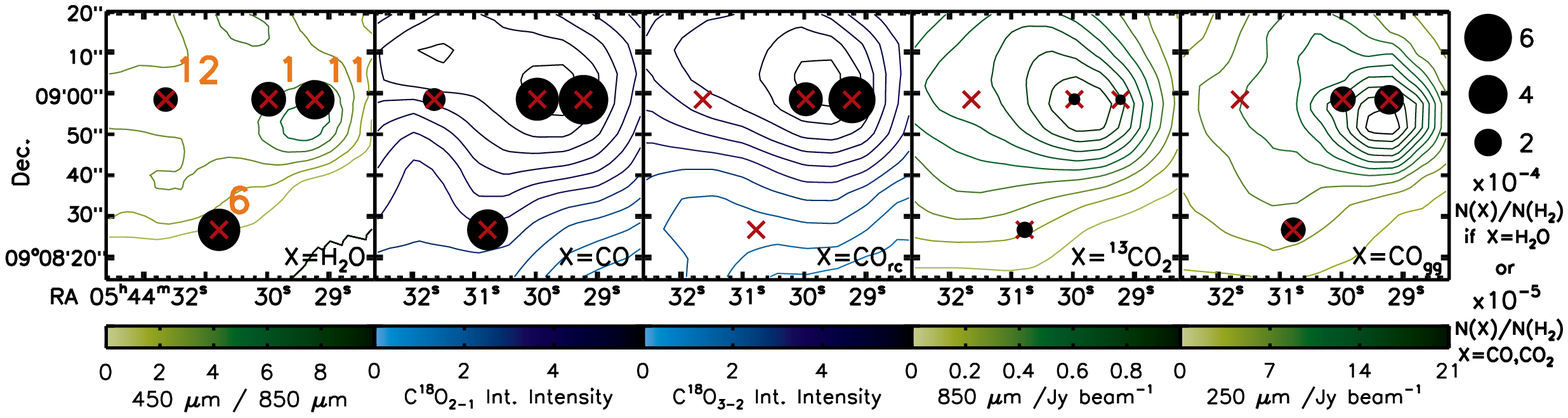}
  \caption{Dust-ice and gas-ice maps of B~35A. Panel 1: 450/850~$\mu$m SCUBA contour map \citep{DiFrancesco08}, with the H$_2$O abundances (H$_2$O/H$_2$) for each object overplotted as black circles; Panel 2: IRAM/HERA C$^{18}$O$_{2-1}$ gas phase contour map \citep{Craigon15} overplotted with CO/H$_2$ abundances; Panel 3: JCMT/HARP C$^{18}$O$_{3-2}$ gas phase contour map (Observing runs M07BU025 and M08BU027, \citet{Buckle,Craigon15}) overplotted with CO$_{rc}$/H$_2$ abundances; Panel 4: SCUBA map at 850~$\mu$m \citep{DiFrancesco08} overplotted with $^{13}$CO$_2$/H$_2$ abundances; Panel 5: {\it Herschel}/SPIRE map at 250~$\mu$m (ObsID 1342227710, \citet{Griffin10}) overplotted with CO$_{gg}$/H$_2$ abundances. It should be noted that the abundances of H$_2$O (Panel 1) are presented in units of 10$^{-4}$, while all other ice maps have units of 10$^{-5}$ (as shown in the abundance scale on the right hand side), and that these abundances are derived from a H$_2$ gas phase map derived from CO observations \citep{Craigon15}.\label{FigScuba3}}
\end{figure*}

One of the original aims of this work was to compare our ice maps to gas phase maps of the same clouds, and investigate the relationships between the solid phase and gas density, temperature, chemistry, and astrophysics. An example of such solid-gas mapping for the core B~35A is presented in Figure~\ref{FigScuba3}, combining our ice data, gas phase C$^{18}$O maps \citep{Craigon15} and archival SCUBA \citep{DiFrancesco08} and {\it Herschel}/SPIRE \citep{Griffin10} data. A contour map of the core constructed from SCUBA maps at 450 and 850~$\mu$m is shown in Panel 1. The 450/850~$\mu$m ratio is related to both the dust temperature and its opacity \citep{Hatchell13}. As the temperature increases through grain heating processes, or the opacity index $\beta$ increases due to grain growth \citep{Martin12}, the 450/850~$\mu$m ratio increases. In B~35A, the highest values are obtained just south of Objects 1 and 12. H$_2$O abundances towards all four objects, calculated from our H$_2$O column densities and H$_2$ gas phase maps of the region \citep{Craigon15}, are overplotted on Panel 1. If the temperature and density are constant across the core, an increase in the 450/850~$\mu$m ratio is likely related to coagulating grains or icy mantles \citep{Ossenkopf94}, and therefore at the highest ratios we'd expect to observe the highest ice abundances. This is true for Objects 12, 1 and 11, but not for Object 6. This indicates that, were we to do quantitative ice mapping, we might be able to see whether ice abundances can establish if changes with dust opacity are directly related to the ice via processes such as coagulation.

In Figure~\ref{FigScuba3}, Panel 2 we present the first combined gas-ice map of a single molecular species, comparing the total CO ice abundance to the C$^{18}$O$_{2-1}$ integrated intensity. C$^{18}$O probes the densest regions of the core, and by comparison with Panel 1, we see that it is approximately colocated with the dust emission.  The relationship between solid and gas phase CO is complex, as it relies on the freeze out of CO under certain conditions of temperature and density. We would expect that, if CO freezeout is simply a function of density, there would be higher CO ice abundances where CO gas is the most abundant, but in this case there is no obvious trend in the column density of solid phase CO compared to gas phase CO. As was the case for H$_2$O (Panel 1) the ``more is more'' pattern is observed for three objects, but Object 6 is once again aberrant. In Panel 3, we consider the component analysis of the CO ice band. Here we plot C$^{18}$O$_{3-2}$ integrated intensity contours, which traces more highly excited CO gas but is observed in essentially the same region as the C$^{18}$O$_{1-2}$ in Panel 2. In Panel 3 we also plot the solid CO observed in a water-rich ice environment, CO$_{rc}$. We only observe CO$_{rc}$ towards the embedded YSO Object 1 and Object 11, with only pure CO observed towards Objects 6 and 12. This would seem to confirm that the CO$_{rc}$ is a tracer of warmer ice, as CO migrates into the water ice layer in warmer, denser regions \citep{Collings03,Karssemeijer14,Lauck15}. It is interesting to note that Objects 1 and 6 have almost identical total CO abundances (Panel 2) but very different CO$_{rc}$ abundances, as has previously been observed in studies of CO ice \citep{Pontoppidan03}.

As shown in Figure~\ref{Fig2Maps} above, CO$_2$ ice was observed upon all four lines of sight in this core, with a variability rougly similar to the total CO abundance variability. However, due to the complexities of extracting column densities for the CO$_2$ data \citep{Noble13} we choose instead to map $^{13}$CO$_2$ ice against the 850~$\mu$m SCUBA map in Figure~\ref{FigScuba3}, Panel 4. 
$^{13}$CO$_2$ ice bands were observed towards three objects studied in \citet{Noble13}, and the derived column density upper limits are presented in Table~\ref{Table1}. For this small sample of objects, we obtain $^{12}$C/$^{13}$C ratios of $\sim$~10 -- 15, significantly lower than previous studies (e.g. $\sim$~50 -- 110, \citet{Boogert00}). Although some of the difference can be attributed to the choices of laboratory ice mixtures used to analyse the two datasets, there is most likely an error introduced by the systematic undersampling of $^{12}$CO$_2$ in our data due to the aforementioned column density extraction issues (detailed in \citet{Noble13}).
In this case, the interest is to determine whether the continuum dust emission can be related to isotopic fractionation in the ice, as would be the case if fractionation had a large temperature dependence. However, it seems that there is no direct relationship between the cooler 850~$\mu$m emission map and the distribution of $^{13}$CO$_2$ in the core.

The final solid-gas map is the CO$_{gg}$ component mapped against the 250~$\mu$m SCUBA data mapping the peak dust emissivity (Figure~\ref{FigScuba3}, Panel 5). As CO$_{gg}$ probes the CO directly adsorbed to bare dust grain surfaces, we would expect to see its abundance increase at higher density as there would be a greater surface area of dust upon which CO could condense. Thus, even if there is more CO ice due to greater density-induced freezeout, it may not fully cover the bare grains and thus the CO$_{gg}$ component would increase. However, in this case we see a roughly constant abundance of CO$_{gg}$ across three objects and no CO$_{gg}$ towards Object 12. For all of the gas-solid maps presented here, full analysis requires a lot more knowledge about the astrophysical environment on a local scale in this small core. It is clear from these maps that the dust in the core of B~35A is centred around Objects~1 and 11, explaining the high column densities of H$_2$O, CO$_2$, and CO ice present towards these objects. However, the column densities of all three species are lower towards Object~12 than Object~6, despite the fact that the dust is more abundant towards Object~12. Thus, dust density is not a direct indicator of ice column density.  The interesting thing with these comparisons is that where dust emission, ice absorption, and gas emission originate from may not necessarily be tracing the same spatial scales. In absorption, the matter is probed along the whole line of sight, whereas emission spectra may probe different depths. This is beyond the scope of the current study, but work is ongoing to answer these questions and in subsequent ice maps, this will be an important factor to incorporate.

%

\section{Conclusions}

The ice mapping results presented in this work indicate that the simple adage ``more is more'' can not be applied to astrophysical ices on a local scale. Although globally this trend is observed when ice abundances are averaged across many lines of sight, mapping probes the large spread of data points in 1D correlation plots. Ice mapping, in this study probing resolutions of $\sim$~2200~AU, gives a striking and immediate visual picture of the ``small scale'' variations which can occur within a cloud or even a core, highlighting the fact that there is no one interstellar ice composition. 

Combined solid-gas maps go one step further, providing vital information on the complex interplay between gas, ice, and dust in star forming regions. Overall, the maps presented in this work reiterate that the presence of more dust in a line of sight does not always indicate that there is more ice, and that both dust density and gas density must be considered to ascertain ice growth mechanisms. There is no simple relationship linking dust (or gas) to ice. They also indicate that features which are often identified as tracers of thermal history or ice processing can't necessarily be immediately linked to the local environment. Even the simple approach presented in this paper, of a direct comparison of ice abundances with available dust and gas data without supporting astrophysical modelling, indicates that ice chemistry is being affected by very local astrophysical conditions.
With this technique, we are tracing the chemistry of a region beyond that which can be ascertained from the gas phase alone. 

In particular, ice mapping could be a powerful method to probe the prestellar to protostellar ice transition. If we can identify specific ice constituents present just before the collapse phase we can set the initial chemical conditions for star formation. With additional modelling and observations, this will lead to a better understanding of how the thermal and processing history of a star forming region leads to ice evolution. Ice mapping is a powerful tool, but it is not as trivial, observationally, as gas phase mapping. Ideally, linking both mapping techniques, using telescopes such as JWST or the E-ELT to produce combined solid-gas maps with less confusion and higher sensitivity, will yield more information on the complex interplay between gas, ice, and dust in star forming regions.

\section*{Acknowledgements}

 J.A.N. acknowledges the financial support of the Royal Commission for the Exhibition of 1851, the University of Strathclyde, the Scottish Universities Physics Alliance, and the Japan Society for the Promotion of Science. The research leading to these results has received funding from the European Community's Seventh Framework Programme FP7/2007-2013 under grant agreement No. 238258 (LASSIE). H.J.F. and J.A.N. are grateful to EU funded COST Action CM0805 ``The Chemical Cosmos: Understanding Chemistry in Astronomical Environments'' for a funding contribution towards the final production of this work.

%

\bsp

\label{lastpage}

\end{document}